\documentclass[twocolumn,showpacs,preprintnumbers,amsmath,amssymb,superscriptaddress,prb]{revtex4}

\usepackage{graphicx,amsmath,amsfonts,amssymb}
\usepackage{dcolumn}
\usepackage{bm}
\usepackage{color}

\newcommand{\mbf}{\mathbf}

\begin{document}
\title{Strain-rate and temperature-driven transition in the shear transformation zone characteristics for 2D amorphous solids}

\author{Penghui Cao}
\author{Harold S. Park\footnote{Corresponding author: parkhs@bu.edu}}
\author{Xi Lin\footnote{Corresponding author: linx@bu.edu}}
\affiliation{Department of Mechanical Engineering, Boston University, Boston, MA 02215, USA}

\date{\today}

\begin{abstract}

We couple the recently developed self-learning metabasin escape algorithm, which enables efficient exploration of the potential energy surface (PES), with shear deformation to elucidate strain-rate and temperature effects on the shear transformation zone (STZ) characteristics in two-dimensional amorphous solids.  In doing so, we report a transition in the STZ characteristics that can be obtained through either increasing the temperature, or decreasing the strain rate.  The transition separates regions having two distinct STZ characteristics.  Specifically, at high temperatures and high strain rates, we show that the STZs have characteristics identical to those that emerge from purely strain-driven, athermal quasistatic atomistic calculations.  At lower temperatures and experimentally relevant strain rates, we use the newly coupled PES + shear deformation method to show that the STZs have characteristics identical to those that emerge from a purely thermally activated state.  The specific changes in STZ characteristics that occur in moving from the strain-driven to thermally activated STZ regime include a 33\% increase in STZ size, faster spatial decay of the displacement field, a change in the deformation mechanism inside the STZ from shear to tension, a reduction in the stress needed to nucleate the first STZ, and finally a notable loss in characteristic quadrupolar symmetry of the surrounding elastic matrix that has previously been seen in athermal, quasistatic shear studies of STZs.

\end{abstract}


\maketitle

\section{Introduction}

Because of their unique combination of high strength and moderate toughness, amorphous solids such as metallic glasses have been extensively studied in recent years~\cite{schuhAM2007,chengPMS2011}.  Much of the study has focused on the inelastic deformation mechanisms that accompany yielding, due to the fact that most amorphous solids fail in a catastrophic and brittle fashion without additional strain hardening immediately following yield.  

Perhaps the most important unresolved issue with regard to the deformation of amorphous solids lies in identifying the characteristics of the unit inelastic deformation mechanism, the shear transformation zone (STZ)~\cite{schuhAM2007,chengPMS2011,argonAM1979,maloneyPRE2006,tsamadosPRE2009,tanguyEPJE2006,lemaitrePRE2007,zinkPRB2006,falkPRE1998,rodneyPRL2009,rodneyPRB2009}, which has primarily been done through both athermal quasistatic shear (AQS)~\cite{maloneyPRE2006,tsamadosPRE2009,tanguyEPJE2006,lemaitrePRE2007}, and classical molecular dynamics (MD) simulations~\cite{zinkPRB2006,falkPRE1998}, and more recently potential energy surface (PES) exploration techniques~\cite{rodneyPRL2009,rodneyPRB2009}.  

Despite these many computational studies, a theoretical framework for characterizing the STZs has until recently been unresolved.  Specifically, researchers have identified that two-dimensional (2D) STZs behave analogously to a classical Eshelby inclusion~\cite{eshelbyPTS1957} embedded within a matrix, where the matrix exhibits a quadrupolar deformation symmetry and where the inclusion represents the size of the STZ~\cite{dasguptaPRL2012,maloneyPRE2006}.

These recent studies~\cite{dasguptaPRL2012,maloneyPRE2006} were performed without accounting for strain rate and temperature effects, and therefore it remains unclear what the structure and characteristics of STZs are at experimentally relevant temperatures and experimentally accessible strain rates.  We address these issues in the present work using a combination of shear deformation and a PES exploration algorithm~\cite{caoPRE2012}, and report the finding of two distinct types of STZs in a 2D binary Lennard-Jones (BLJ) solid: those that have identical characteristics to those that emerge from purely strain-driven, athermal quasistatic atomistic calculations, and those that emerge from a purely thermally activated state.  We further show that the transition is characterized by changes in the inclusion size, the matrix deformation symmetry, the localized strain and displacement fields, the local free volume, the deformation mechanism inside the STZ, and finally the STZ nucleation stress.

\section{Numerical Methodology}

\subsection{Self-Learning Metabasin Escape Algorithm}

The self-learning metabasin escape (SLME) algorithm\cite{caoPRE2012} is implemented in this work to explore the PES for two purposes.  First, it corresponds to purely thermal activation of the amorphous solid in the absence of any applied shear deformation, as described in detail in this section.  Second, it also explores the PES at each state of strain for any strain rate, as explained in detail in Secs. IIB below.  

\begin{figure} \begin{center} 
\includegraphics[scale=0.3]{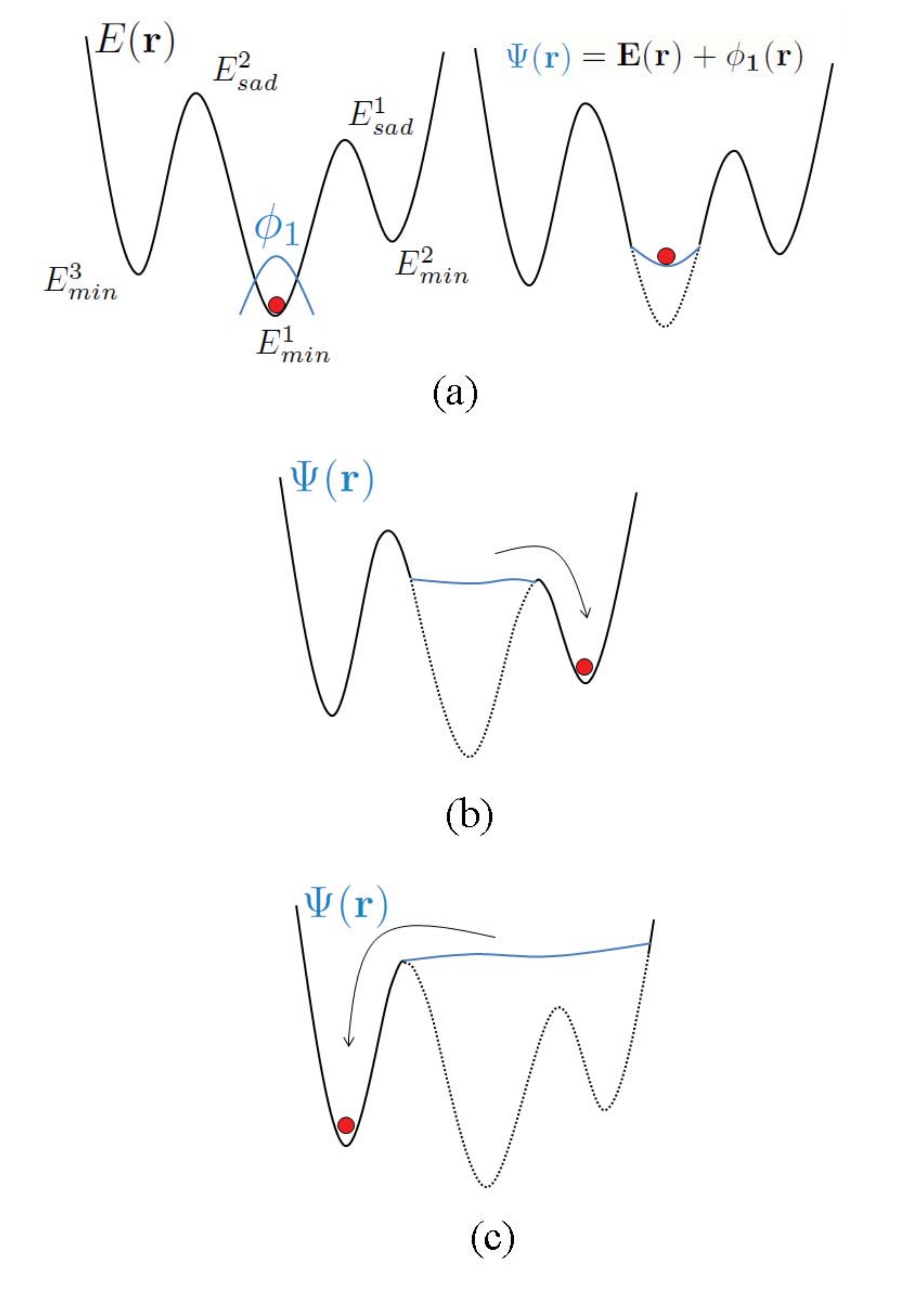}
\caption{\label{}Schematic of the autonomous basin climbing (ABC) potential energy surface exploration technique.  (a) Depiction of how the addition of a penalty function $\phi_{1}$ to the PES defined by $E(\mbf{r})$ results in the penalty function modified PES defined by $\Psi(\mbf{r})$.  (b,c) The addition of more penalty functions results in the system being pushed out via $\Psi(\mbf{r})$ of various local minima into other, neighboring energy basins.  $E_{min}$ and $E_{sad}$ correspond to energy minima and saddle points, respectively.  We emphasize that while the PES depicted in this figure is one-dimensional, the SLME algorithm utilized in this work explores the 2$N$-dimensional PES, where $N$ is the number of atoms in the simulation.}
\label{abc} \end{center} \end{figure}

The SLME algorithm is a self-learning version of the autonomous basin climbing (ABC) algorithm recently developed by ~\citet{kushimaJCP2009}.  The basic ABC algorithm implementation works in an intuitive manner, sketched briefly in Fig. \ref{abc}.  Starting from any initial energy minimum configuration, localized penalty functions $\phi_{i}(\mbf{r})$ are successively applied to assist the system in climbing out of the current local energy well and exploring other, neighboring energy wells. Physically, this corresponds to activation of the system due to thermal effects, and it is how we obtain the characteristics for thermally activated STZs later in this work.  Mathematically, this is written as
\begin{equation}\label{eq:abc1} \Psi(\mbf{r})=E(\mbf{r})+\sum_{i=1}^{p}\phi_{i}(\mbf{r}),
\end{equation}
where $\Psi(\mbf{r})$ is the augmented potential energy due to the addition of the penalty functions, $E(\mbf{r})$ is the original potential energy function, i.e., the BLJ potential in the present case, and $p$ is the total number of penalty functions.  Although in principle any type of localized functions (i.e., Gaussians~\cite{kushimaJCP2009}, ~\cite{kushimaJCP2009_2}) can be used in Eq. (\ref{eq:abc1}), we chose quartic penalty functions in this work due to their desirable property of naturally vanishing energy and forces at the penalized subspace boundaries \cite{caoPRE2012}.

As can be inferred from Eq. (\ref{eq:abc1}), many small penalty functions are needed in order to push the system out of a given energy basin.  All of these penalty functions must be kept such that the system does not fall back into an energy basin that has already been explored.  Clearly, the requirement to store all previous penalty functions becomes prohibitive as more and more energy basins are explored.  Because of this, the computational expense associated with the ABC method increases substantially ~\cite{caoPRE2012}, and becomes the bottleneck of the ABC method, as the PES exploration continues.  

This issue was alleviated substantially in the SLME algorithm developed recently by ~\citet{caoPRE2012}, where a few self-learning schemes were introduced.  The essential idea is that instead of storing all of the (many) penalty functions that have been used to boost the system out of the different energy wells it has explored, the penalty functions are self-updated in various ways such that, upon exiting a given energy well, only a few independent ones remain.  Specifically, as the system evolves on the PES, new penalty functions can be self-generated according to the history without any preassumed parameters.  These newly self-generated penalty functions and all the previously imposed penalty functions are then subject to iterative reconstructions to minimize their ($2N+1$)-dimensional spatial overlap in the penalized configurational subspace, where $N$ is the total number of atoms in the system.  Therefore, redundant penalty functions can be identified and removed effectively as the system evolves, so that the total amount of penalty functions can be reduced to a minimal amount.  In addition, such self-learned reconstructions offer new flexibility to the penalty functions, so that they can  self-adapt to the underlying energy landscape and their size distributions can naturally reflect the actual sizes of metabasins. This approach, called the SLME algorithm, together with the resulting decrease in penalty function storage requirements, was shown to lead to an substantial increase in computational efficiency as compared to the previous ABC implementation~\citep{caoPRE2012}.

Thus, by repeating the alternating sequence of penalty function addition and augmented energy relaxation, the system is self-activated to fill up the local energy basin and escape through the lowest saddle point.  By maintaining all the independent penalty functions imposed during the SLME trajectory, frequent recrossing of small barriers is eliminated, which is a significant advantage of such history-penalized methods~\citep{laioPNAS2002,kushimaJCP2009,caoPRE2012}.  We emphasize that while Fig. \ref{abc} depicts the ABC method in one dimensional, in actuality for the present work the SLME approach investigates the entire, 2$N$-dimensional (2$N$-D) PES, where $N$ is the total number of atoms in the system.

There are a few major advantages of using the SLME method as compared to other PES exploration techniques.  For example, the SLME approach does not need to specify the softest eigenmode searching direction as in hyperdynamics~\citep{voterPRL1997} or dimer methods~\citep{henkelmanJCP1999}, or to restrict the searching subspace as in metadynamics~\citep{laioPNAS2002}.  It is also relevant to discuss this approach in contrast to nudged elastic band (NEB) techniques that have recently been utilized to study the deformation mechanisms of nanostructured metals at experimentally-relevant time scales~\citep{zhuPRL2008}.  The NEB approach is particularly well suited for metal plasticity because it requires {\it a priori} knowledge of the final configuration in order to find the minimum energy pathway.  In the case of metals, it is well known that crystal defects such as twins, dislocations, and stacking faults are the likely plastic deformation mechanisms~\citep{parkJMPS2006}.  However, the situation is quite different for amorphous solids, where the atomic structure of the equivalent basic deformation mechanism, the STZ, remains unknown~\citep{schuhAM2007,chengPMS2011}.  

As a brief summary, the SLME approach discussed above resolves two critical issues with regard to mapping out the PES of amorphous solids.  First, the SLME algorithm~\citep{caoPRE2012} is the only demonstrated computational approach that can systematically explore sequential metabasin activation events on the complete 2$N$-D PES without {\it a priori} knowledge of final states or order parameters as required for supercooled liquids and amorphous solids.  Second, the computational efficiency of the SLME approach is critical as it enables us to get access to a sufficiently large configurational space by infrequent free-energy activation events over very large activation free energy barriers $Q^{*}(T,\dot{\gamma})$ that are needed to access temperatures $T$ well below the glass transition temperature and at the laboratory shear strain rates $\dot{\gamma}$. In the following section, we explain how to incorporate temperature and strain-rate effects.
 
\subsection{Shear-Coupled Self-Learning Metabasin Escape Algorithm}

\subsubsection{Temperature and strain-rate dependent activation free energy $Q^*(T,\dot{\gamma})$ formalism}

To incorporate the effects of strain rate and temperature, we begin with the following expression for a single-event shear strain rate $\dot{\gamma}_{\rm single}$, which is derived~\citep{zhuPRL2008} from the transition state theory for constant temperature. The gives the most likely nucleation rate for STZs:
\begin{equation}\label{eq:tst3}
\dot{\gamma}_{\rm single}=nv_0\dfrac{k_{\rm B} T}{\mu \Omega}\exp\left[-\dfrac{Q(T)-TS_{\rm c}}{k_{\rm B} T}\right],
\end{equation}
where $n$ is the number of independent STZ nucleation sites, $v_0$ is the attempt frequency, $\mu$ is the shear modulus, $\Omega$ is the activation volume, and $S_{\rm c}$ is the activation configurational entropy that is primarily due to anharmonic thermal expansion and thermal softening effects~\citep{ryuPNAS2011}, which are partially captured during the initial slow quenching stage needed to obtain the initial configuration by the stress-free NPT (constant number of
particles, and constant pressure and temperature) ensemble.

It can be seen in Eq. (\ref{eq:tst3}) that the single-event nucleation rates $\dot{\gamma}_{\rm single}$ do not necessarily follow the Arrhenius relation since the activation energy $Q(T)$ can be strongly dependent on temperature, which is one of the well known phenomena occurring in many relaxation events of supercooled liquids and amorphous solids. \cite{kushimaJCP2009}  Such strong temperature dependence is inherited directly from the free energy $F_{i}(T)$ of the $i$th local minimum basin~\citep{liPO2011} as $Q_{ij}(T)= Q_{ij}[F_i(T),F_j(T)]$.  Namely, the temperature dependence of $Q_{ij}(T)$ has a {\it functional} dependence through the temperature dependent free energy of the initial and final inherent structures, $F_i(T)$ and $F_j(T)$, respectively.  While the SLME trajectories are along the PES, the detailed balance between any two thermally equilibrated free energy basins $F_{i}(T)$ and $F_{j}(T)$ is enforced at all times by the standard Monte Carlo method as to be discussed below.  Therefore, the entire collection of \{$F_i(T), Q_{ij}(T)$\} forms an ergodic system with only Markov chain transitions being allowed.  One may refer to~\citet{liPO2011} and~\citet{kushimaJCP2009} for more detailed discussions.

\begin{figure} \begin{center} 
\includegraphics[scale=0.35]{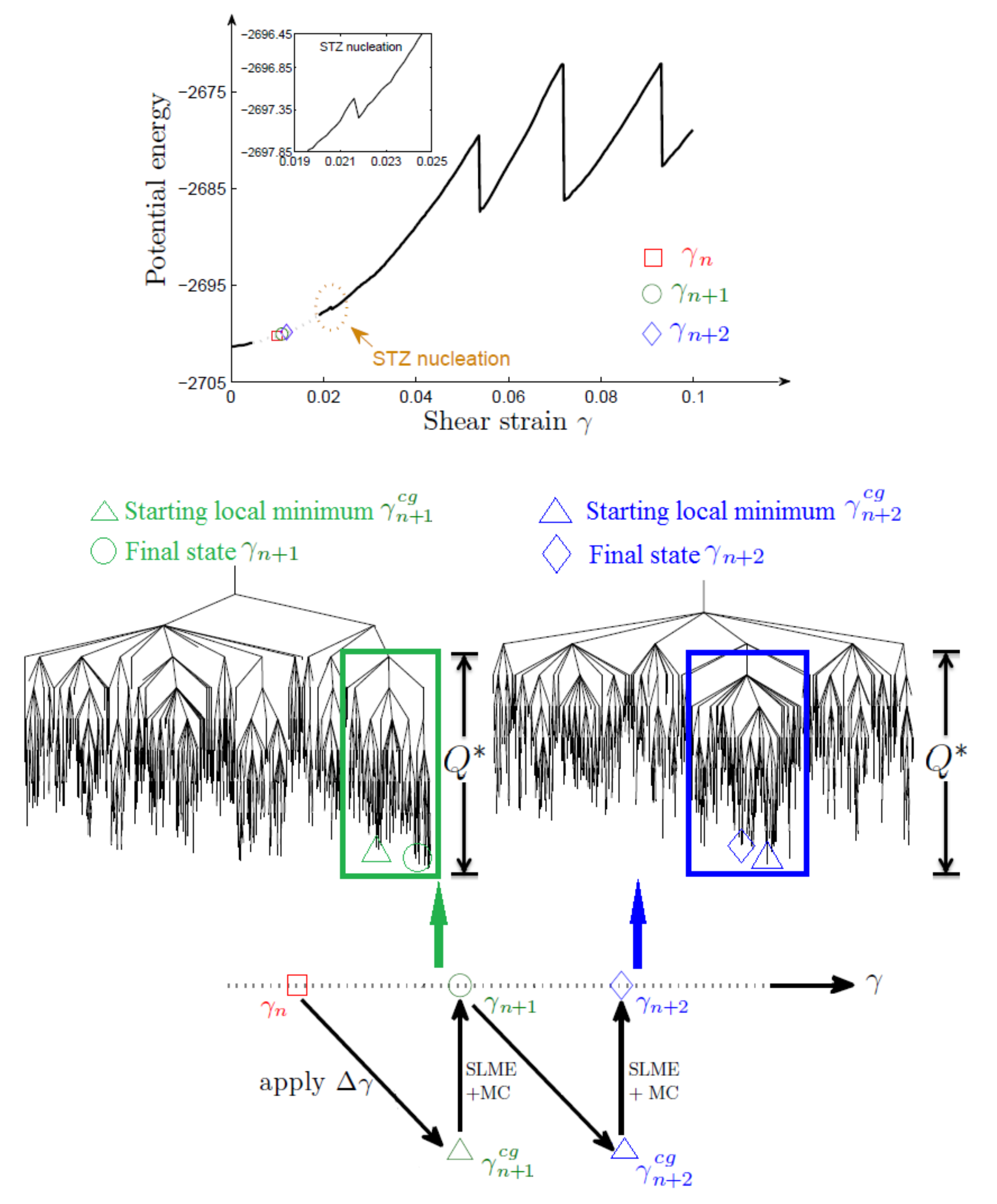}
\caption{\label{}Illustration of how the SLME method~\citep{caoPRE2012} is utilized to find the equilibrium configuration after a given strain increment $\Delta\gamma$ is applied to the system.  Specifically, starting from $\gamma_{n}$, a shear increment $\Delta\gamma$ is applied to the system.  At that point, a standard conjugate gradient (CG) energy minimization is performed while keeping the strain fixed, giving the state $\gamma_{n+1}^{cg}$.  Starting from the energy minimized configuration $\gamma_{n+1}^{cg}$, the SLME method is used to determine the potential energy tree structure as shown, where the lower end point of each vertical line specifies an independent local minimum energy configuration, and where every pair of these local minima is connected by a unique saddle point specifying the lowest activation energy barrier between them.  We truncate the tree structure to only enable energy transitions below $Q^{*}$, as shown in the green box.  Finally, a classical Monte Carlo algorithm is employed to find, among the hundreds of local minima in the green box, the most likely equilibrium configuration, which is then denoted $\gamma_{n+1}$. The same procedure is then utilized to find the next equilibrium configuration for the strain $\gamma_{n+2}$, though we note that the PES tree structure is different at the new shear strain $\gamma_{n+2}$, which again is mapped out using the SLME method.}
\label{fig1a} \end{center} \end{figure}

Following this new formalism, we need to extend the \emph{single-event} activation energy $Q(T)$ to the temperature- and strain-rate-dependent \emph{many-event} $Q^*(T,\dot{\gamma})$, where $Q^{*}(T,\dot{\gamma})$ contains many (hundreds) of such activation events, as illustrated by thegreen box in Fig. \ref{fig1a}.  Specifically, $Q^{*}(T,\dot{\gamma})$  is the maximal activation energy with respect to the initial free energy basin $F(T)$, so that $Q^*(T,\dot{\gamma})$ truncates the ergodic Markovian system into an ergodic Markovian subspace and the remainder, the part that is not accessible at the given strain rate $\dot{\gamma}$.  As $\dot{\gamma}$ decreases, the ergodic Markovian subspace increases monotonically, with the important implication that more and more mechanical deformation pathways that were not accessible at high strain rates become accessible assuming that the PES exploration technique (i.e. the SLME approach) is able to reach and climb over the corresponding energy barriers on the PES.  Because the SLME approach enables us to efficiently access and calculate the allowed activated states $Q(T,\dot{\gamma}) \le Q^*(T,\dot{\gamma})$ for essentially arbitrarily large $Q^{*}(T,\dot{\gamma})$, we are able to compute the yield stress $\tau (T, \dot{\gamma})$ and activation volume $\Omega(T,\dot{\gamma})$ at all relevant temperatures and shear strain rates $\dot{\gamma}$ ranging from MD to experimentally accessible.

Having established the theoretical basis for extending $Q(T)$ to $Q^{*}(T,\dot{\gamma})$, we note that, in contrast to the simpler deformation processes occurring in crystalline materials~\citep{zhuPRL2008}, the coupled thermomechanical deformation events in amorphous solids likely consist of multiple sequential activation events.  

Therefore, by defining a characteristic prefactor from Eq. (\ref{eq:tst3}) as $\dot{\gamma_0}= \dfrac{k_{\rm B} Tnv_0}{\mu \Omega}\exp\left(\dfrac{S_{\rm c}}{k_{\rm B}}\right)$ that is known to be weakly $T$ dependent \citep{johnsonPRL2005,chengAM2011}, we can rewrite Eq. (\ref{eq:tst3}) as
\begin{equation}\label{eq:tst3a}
\dot{\gamma}_{\rm single}=\dot{\gamma_{0}}\exp\left[-\dfrac{Q(T)}{k_{\rm B} T}\right].
\end{equation}
Finally, by converting from $Q(T)$ to $Q^{*}(T,\dot{\gamma})$ based on the above discussion, we can construct the maximal activation energy barrier by rearranging Eq. (\ref{eq:tst3a}) as
\begin{eqnarray}\label{eq:tst4}
Q^*(T,\dot{\gamma})= -k_{\rm B} T \ln\left(\dfrac{\dot{\gamma}}{\dot{\gamma_{0}}}\right),
\end{eqnarray}
which defines the ergodic Markovian region in the entire SLME connectivity tree structures at the given strain rate, for example as shown as the green box in Fig. \ref{fig1a}.  Within this ergodic window, all the transitions follow the Markov chain processes to reach a local equilibrium, so that the amorphous solid (BLJ) system can relax to the accessible lowest free-energy configurations.

\subsubsection{Algorithmic Details}

We now detail how the SLME method is coupled with shear deformation and the classical Monte Carlo to calculate the stress and equilibrium atomic positions of the BLJ solid as a function of strain, strain rate, and temperature, with no change in methodology needed to distinguish between elastic and plastic strain increments.  After obtaining the initial stress-free glassy structures for a given temperature, we apply the following algorithm for all loading increments.  Specifically, assume that, as shown in Fig. \ref{fig1a}, the system exists at shear strain $\gamma_{n}$.  We then apply a shear strain increment $\Delta\gamma=0.01\%$, followed by a standard conjugate gradient energy minimization to find the resulting equilibrium positions of the atoms, which brings us to the shear strain state $\gamma_{n+1}^{cg}$ in Fig. \ref{fig1a}.  It is important to note that the system size and boundaries are held fixed during the energy minimization such that the shear strain $\gamma_{n+1}^{cg}=\gamma_{n}+\Delta\gamma$.

From that point, the SLME approach~\citep{caoPRE2012,kushimaJCP2009} is utilized to explore the PES at the strain $\gamma_{n+1}^{cg}$, as illustrated via the potential energy connectivity tree structures~\citep{wales2003} shown in Fig. \ref{fig1a}, while again the system size and boundaries are held fixed.   Importantly, we only allow transitions within the SLME connectivity tree structures below the maximum energy barrier $Q^{*}$ shown in Eq. (\ref{eq:tst4}) as highlighted by the green box shown in Fig. \ref{fig1a}.  The maximum energy barrier $Q^{*}$ is a defined parameter that specifies the maximum barrier height on the PES that can be overcome, via thermal assistance, for a given strain rate.  This is because in physical terms, choosing a value of $Q^{*}$ is equivalent, as shown in Eq. (\ref{eq:tst4}), to specifying the strain rate of the simulation for a given temperature.  In other words, for very high strain rates as seen in MD simulations, only small energetic barriers $Q^{*}$ can be crossed for each strain increment due to the small amount of time given to the system to explore other possible, thermally-assisted configurations.  In contrast, at slower strain rates, the system has more time between successive strain increments such that it can explore many other possible, thermally-assisted configurations, and thus potentially climb over larger energy barriers, with the sole restriction that the thermally assisted barrier crossing must be smaller than $Q^{*}$.  It is important to note, however, that we do not enforce that the maximum barrier height $Q^{*}$ is crossed for each strain increment.

Summarized a different way, the picture of deformation underlying our work is one that receives contributions due to both mechanical and thermal work.  The mechanical work dominates the deformation process at high strain rates, when the time in between strain increments is not sufficient to enable substantial, thermally-assisted atomic motion.  Thermal work is viewed as making a substantial contribution to the deformation process at slower strain rates, when sufficient time to enable thermally-driven deformation in between successive strain increments is provided to the system.

As shown in Fig. \ref{fig1a} starting from the specific strain state $\gamma_{n+1}^{cg}$, the SLME algorithm typically finds on the order of a few hundred local minima for each value of shear strain, which gives on the order of ten thousand local minima for the entire shear deformation process, as well as all of the corresponding lowest energy barriers between every pair of these local minima.  In other words, at a given strain rate, the system can self-explore the PES via the SLME approach by climbing over all the allowed energy barriers that are smaller than $Q^{*}$, as depicted via the green boxed portion of the PES connectivity tree structure in Fig. \ref{fig1a}.  Within this truncated potential energy subspace, we identify the most likely free-energy basin, namely the basin with the lowest free energy at this instantaneous NVT (constant number of particles, and constant volume and temperature) ensemble at the given strain state, via the standard Monte Carlo method.  This lowest free-energy basin at strain $\gamma_{n+1}=\gamma_{n+1}^{cg}=\gamma_{n}+\Delta\gamma$, as shown by the green circle in Fig. \ref{fig1a}, is assigned to be the initial configuration for the next loading increment.  Furthermore, the atomic configuration corresponding to the lowest free-energy basin corresponds to the shear strain state $\gamma_{n+1}$.  The shear stress corresponding to the shear strain $\gamma_{n+1}$ is then obtained by calculating the virial stress based upon the atomistic configuration at $\gamma_{n+1}$.  At this point, a new shear strain increment of 0.01\% is applied and the SLME process as just described is repeated until the yield stress is obtained, where the yield stress is determined to be the maximum stress that is reached before the first substantial stress drop signifying yield is obtained.  

\section{Results}

The 2D BLJ solid of ~\citet{falkPRE1998} we considered in this work contained $N=1000$ particles of the same unit mass under periodic boundary conditions. The material contained two types of particles, with a large-to-small particle ratio of 447:553. The glass was prepared by quenching from a liquid state~\cite{falkPRE1998} to well below the glass transition temperature $T_g=0.3$ in a constant volume ensemble with a cooling rate of $2\times10^{-7}$. After quenching, the amorphous structures were relaxed to zero average stress states using an NPT ensemble.  For the BLJ potential $\sigma_{SL}$ and $\epsilon_{SL}$ have units of length and energy, respectively, while the mass of all particles is $m=1$.  With these defined, the reduced time is written as $t_{0}=\sigma_{SL}\sqrt{m/\epsilon_{SL}}$, the shear modulus is $\epsilon_{SL}/\sigma_{SL}^{2}$, while the strain rate $\dot{\gamma}$ units are ($\sigma_{SL}\sqrt{m/\epsilon_{SL}})^{-1}$.  All units in this paper are given in reduced LJ form, where a comprehensive description of LJ reduced units is given in Appendix B of Ref. ~\citet{allen1987}.  Finally, we should also emphasize that our choice of studying a 2D and not 3D amorphous solid using the aforementioned BLJ potential does not preclude the formation and propagation of localization instabilities in two dimensional such as shear bands~\cite{falkPRE1998,maloneyPRE2006,tsamadosPRE2009}.

\subsection{Defining the Characteristics of Two-Dimensional Strain-Driven and Thermally-Activated Shear Transformation Zones}

Before assessing the coupled effects of strain rate and temperature on the STZ characteristics, we first discuss and define the STZ characteristics for two limiting cases.  In the first case, the system was first quenched to a temperature of $T=0.001T_{g}$.  Shear strain increments of 0.01\% were subsequently applied to the quenched structure followed by conjugate gradient energy minimization, where this scenario corresponds to the limiting case of purely shear strain ($\gamma$)-driven deformation at very low, or effectively zero, temperature. This scenario is typically called athermal quasistatic shear (AQS) in the literature~\cite{maloneyPRE2006,dasguptaPRL2012}, and we keep that nomenclature here.  

The second case corresponds to purely thermal activation of the system in the absence of any shear deformation using the previously discussed SLME algorithm~\cite{caoPRE2012}.  Specifically, the initial stress-free configuration after quenching is activated by imposing $(2N+1)$-D history-based penalty functions followed by energy minimization. A series of activation and relaxation steps can make the system escape from the current basin and move to a neighboring energy well. The SLME trajectories consist of free energies of all the inherent structures $F_i(T)= F({\bf S}_i; T)$ that have been visited, where ${\bf S}_i$ are the 2$N$-D position vectors of local minima, as well as the  activation free energies $Q_{ij}(T)$ between all pairs.  Following these trajectories, we can activate the system to cross sufficiently large energy barriers that cause strain localization~\cite{rodneyPRL2009} via pure thermal activation, in the absence of any applied shear deformation.

We focus in this work on the first plastic event, i.e., the development of the initial STZ, rather than on the structure of the STZ at yield or on the nature of STZ interactions leading to failure via shearbanding.  This first plastic event is identified by a small drop in the potential energy versus strain curve, and occurs in the AQS simulation at 3.1\% shear strain, which is well below the yield strain of 7.1\%.  Unlike in centrosymmetric crystalline solids, the forces acting on atoms in an amorphous solid are nonzero after a small homogeneous strain increment from an equilibrium state due to the lack of crystalline symmetry, and a nonaffine displacement $\delta{\bf u}$ is necessary to bring the system to a local energy minimum.  Here $\delta{\mbf{u}}$ is defined as  
\begin{eqnarray}
\delta{\bf {u}}= \bf{u} - \bf {u}^{born},
\end{eqnarray} 
where $\bf u$ is the displacement and the Born term $\bf{u}^{born}$ corresponds to the homogeneous contribution to the displacement $\mbf{u}$.  Similarly, we can define a nonaffine local strain $\delta{\eta}$ as

\begin{eqnarray}
\delta{\bf{\eta}} = \bf{\eta} - \bf{\eta}^{born},
\end{eqnarray} 

where the von Mises strain $\eta$ is obtained following~\citet{falkPRE1998} and $\eta^{\rm born}$ is the applied strain on the simulation box.  Thus, while the von Mises strain $\eta$ is always positive, the nonaffine local strain $\delta\eta$ can be negative as the strain applied to the simulation box can be larger than the strain on an individual atom.

\begin{figure} \begin{center} 
\includegraphics[scale=0.55]{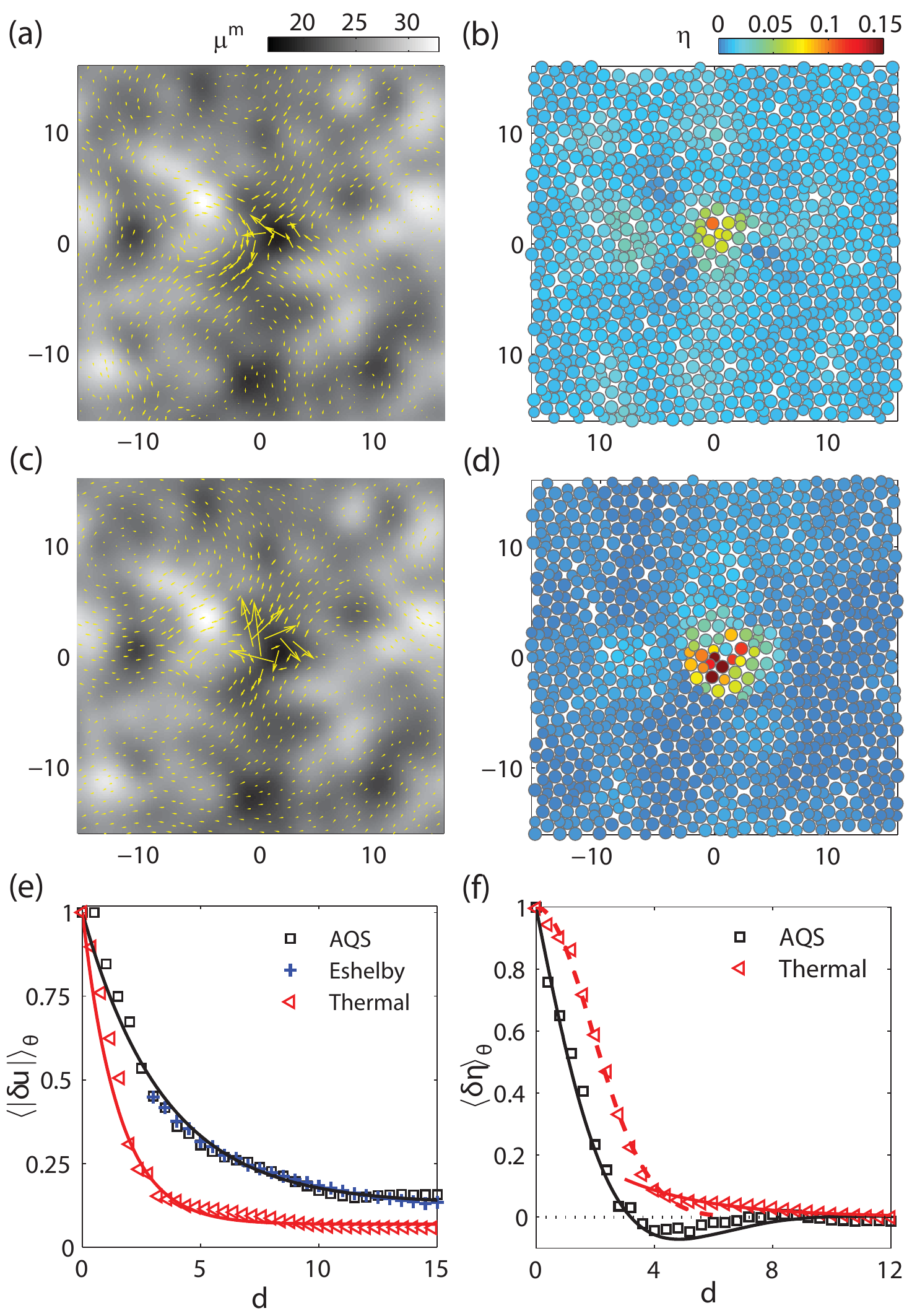}
\caption{(Color online) Nonaffine displacement $\delta\bf{u}$ at STZ nucleation and $T=0.001T_{\rm g}$ computed by (a) AQS and (c) purely thermal activation, with the background colored by the local shear modulus~\cite{YoshimotoPRL2004,lutskoJAP1988} $\mu^{m}$.  Von Mises local shear strain $\eta$ at STZ nucleation and $T=0.001T_{\rm g}$ for (b) AQS and (d) purely thermal activation.  Angularly averaged nonaffine (e) displacement magnitude $\left<|\delta{\bf u}|\right>_{\theta}$ and (f) local strain $\left<\delta{\eta} \right>_{\theta}$ as a function of distance $d$ from the center of the STZ.  Both black (upper) and red solid lines in (e) decay exponentially, while the black  and red lines (upper) in (f) follow Eqs. (\ref{eq:decay2}) and (\ref{eq:decay3}), respectively.}  
\label{fig1} \end{center} \end{figure}

Figures \ref{fig1}(a) and (c) show the total nonaffine displacement field for AQS and thermal activation (calculated using SLME~\cite{caoPRE2012}), respectively, at the formation of the first STZ, with the background colored by the local shear modulus $\mu^{m}$~\cite{YoshimotoPRL2004,lutskoJAP1988}.  $\mu^{m}$ was calculated by partitioning the simulation box into small squares of length $L/10$, with $L$ being the simulation box length.  The angularly averaged nonaffine displacement magnitudes $\left<|\delta{\bf u}|\right>_{\theta}$ are plotted in Fig. \ref{fig1}(e) as a function of the distance $d$ from the center of the STZ.  It is clear that the purely strain-driven AQS results (black squares) agree well with the analytic solution based on the Eshelby inclusion (blue pluses) as expected~\cite{dasguptaPRL2012}.  They both follow the same exponential decay function $\left<|\delta{\bf u}|\right>_{\theta}= (1- \delta u _{\infty})\exp(-kd) + \delta u_{\infty}$, where for AQS the decay exponent $k= 0.29$ and the far-field nonaffine displacement plateau $\delta u_{\infty}= 0.13$.  In contrast, the nonaffine displacement field for the purely thermally activated STZ (red triangles) not only decays significantly faster with $k= 0.66$, but also decays to a much smaller far-field plateau $\delta u_{\infty}= 0.068$.  This substantially lower far-field nonaffine deformation is related to the fact that systems containing thermally activated STZs are able to reduce the strain energy throughout the amorphous system by localizing the plastic deformation more effectively than in the AQS cases.

In addition, the local von Mises shear strain $\eta$ offers a complementary view of these self-localized STZs, as summarized in Figs. \ref{fig1}(b) and (d) for the AQS and purely thermal cases, respectively. The angularly averaged nonaffine local strain $\left<{\delta{\eta}}\right>_{\theta}$ as a function of distance $d$ from the STZ center is plotted in Fig. \ref{fig1}(f). Interestingly, the AQS strain field [Fig. \ref{fig1}(f), black squares] does not decay monotonically as the AQS nonaffine displacement field [Fig. \ref{fig1}(e)].  Instead, it contains nontrivial oscillations within the overall exponentially decay profile as
\begin{eqnarray}\label{eq:decay2}
\left<\delta{\eta}\right>_{\theta}= \cos(k_{\rm o} d)\exp(-k_{\rm d} d) ,
\end{eqnarray}
where the oscillative wave vector $k_{\rm o}= 0.49$ and the decay exponent $k_{\rm d}= 0.47$.  Furthermore, we define the strain-driven STZ core size as $d_{\rm c}= \pi/2 k_{\rm o}= 3.2$, indicating the matrix relaxation immediately surrounding the STZ.

In sharp contrast, the local strain field of the purely thermally activated configuration [Fig. \ref{fig1}(f], red triangles) has a random-walk Gaussian core overlapped by a far-field exponentially decaying tail:   
\begin{align}\label{eq:decay3}
   \left<\delta\eta\right>_{\theta}  =
        & 
    \left\{ \begin{array}{l}
         \exp(-d^2/2\sigma^2),  d \le d_{\rm c}\\
         \exp\left[-k\left(d-d_{\rm c}\right)-d_{\rm c}^2/2\sigma^2 \right], d\ge d_{\rm c}
    \end{array}\right., 
\end{align}
where the Gaussian variance $\sigma= 1.89$, the exponential decay rate $k= 0.34$, and the STZ core distance $d_{\rm c}= 4.3$.  This purely thermal STZ core contains about 56 atoms, as compared to 31 atoms for the AQS case.  As a summary of Figs. \ref{fig3}(a)-(f), we note that the thermally activated STZs have significantly larger core areas and faster exponential decay rates than the purely strain-driven STZs.

Having established the two distinct STZ characteristics, it is important to address why the two types of STZs have different decay lengths in the strain and displacement fields, and why the thermally assisted STZs are larger than the strain-driven STZs.  This is because the thermally assisted STZs are able to grow in size in comparison to strain-driven STZs by diffusive processes.  Furthermore, this also provides a mechanism for the matrix surrounding the STZ to reach lower energy, less strained configurations than for the matrix surrounding the STZ in the strain-driven case.  Finally, this explains why the decay rate for the displacement fields surrounding thermally assisted STZs is much faster than for the displacement fields surrounding the purely strain-driven STZs.

\begin{figure} \begin{center} 
\includegraphics[scale=0.55]{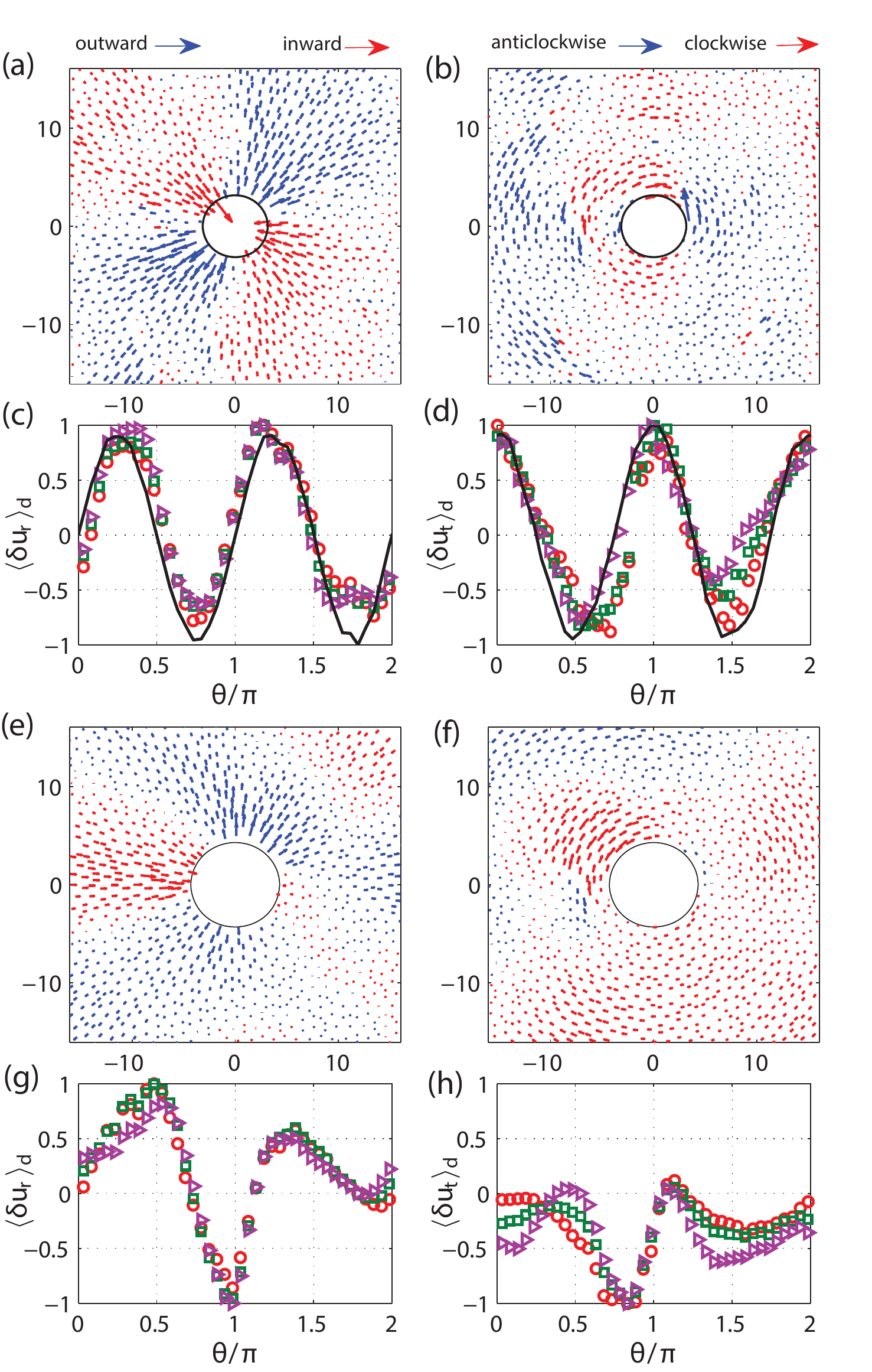}
\caption{ (Color online)(a) Radial $(\delta\mbf{u}_{\rm r})$ and (b) tangential $(\delta\mbf{u}_{\rm t})$ projections of the AQS nonaffine displacement field $\delta\mbf{u}$ [Fig. \ref{fig1}a] at STZ nucleation and $T=0.001T_{g}$. (a, b) The black circles are $d_{\rm c}= 3.2$. (c, d) The normalized angle-resolved $|\delta{\bf  u}_{\rm r}(\theta)|$ and  $|\delta{\bf  u}_{\rm t}(\theta)|$ shown at $d \le 9$ (red circles), 12 (green squares), and 15 (purple triangles), respectively.  (e-h) The purely thermal activation [Fig. \ref{fig1}c] results, in the same order as (a-d).  The black circles in (e) and (f) are $d_{\rm c}= 4.3$.  Black lines in (c) and (d) are the Eshelby inclusion results at $d= 15$.}
\label{fig2} \end{center} \end{figure}

In Figs. \ref{fig2}(a) and (b), we show the radial $(\delta\mbf{u}_{\rm r})$ and tangential $(\delta\mbf{u}_{\rm t})$ components of the nonaffine displacement field outside of the AQS core radius $d_{\rm c}= 3.2$, and their angle-resolved magnitudes in Figs. \ref{fig2}(c) and (d).  It is important to emphasize that the non-affine displacements in Figs. \ref{fig2}(a) and (b) correspond to the deformation of the matrix surrounding the STZ (inclusion) which has been centered in the middle of the image, where an empty hole of radius $d_{\rm c}=3.2$ has been drawn to represent the STZ core as previously discussed.  The quadrupolar deformation of the matrix surrounding the central STZ core region is clearly present in these AQS results, which agrees well with the analytic Eshelby solutions~\cite{dasguptaPRL2012} [Figs. \ref{fig2}(c) and (d), black lines].  By fitting to our numerical results to Eq. (8) of Ref. ~\citet{dasguptaPRL2012}, we obtain the inclusion radius $a=3.2$, shear eigenstrain $\epsilon^{*}=8.0\%$, and Poisson's ratio $\nu = 0.32$.  However, our thermal activation simulation results in Figs. \ref{fig2}(e) and (f) do not exhibit quadrupolar symmetry, which is in agreement with other recent PES results~\cite{rodneyPRL2009}.  For this purely thermal activation study using the SLME method, more than 200 sequential activated events were sampled, and 90\% in absence of quadrupolar symmetry.  The snapshots in Figs. \ref{fig2}(e) and (f) correspond to the minimum energy state obtained following the crossing of the highest saddle point, which results in a large local strain in the STZ core exceeding 10\%. 

\subsection{Strain Rate and Temperature Effects on Two-Dimensional STZ Characteristics}
  
\begin{figure} \begin{center} 
\includegraphics[scale=0.58]{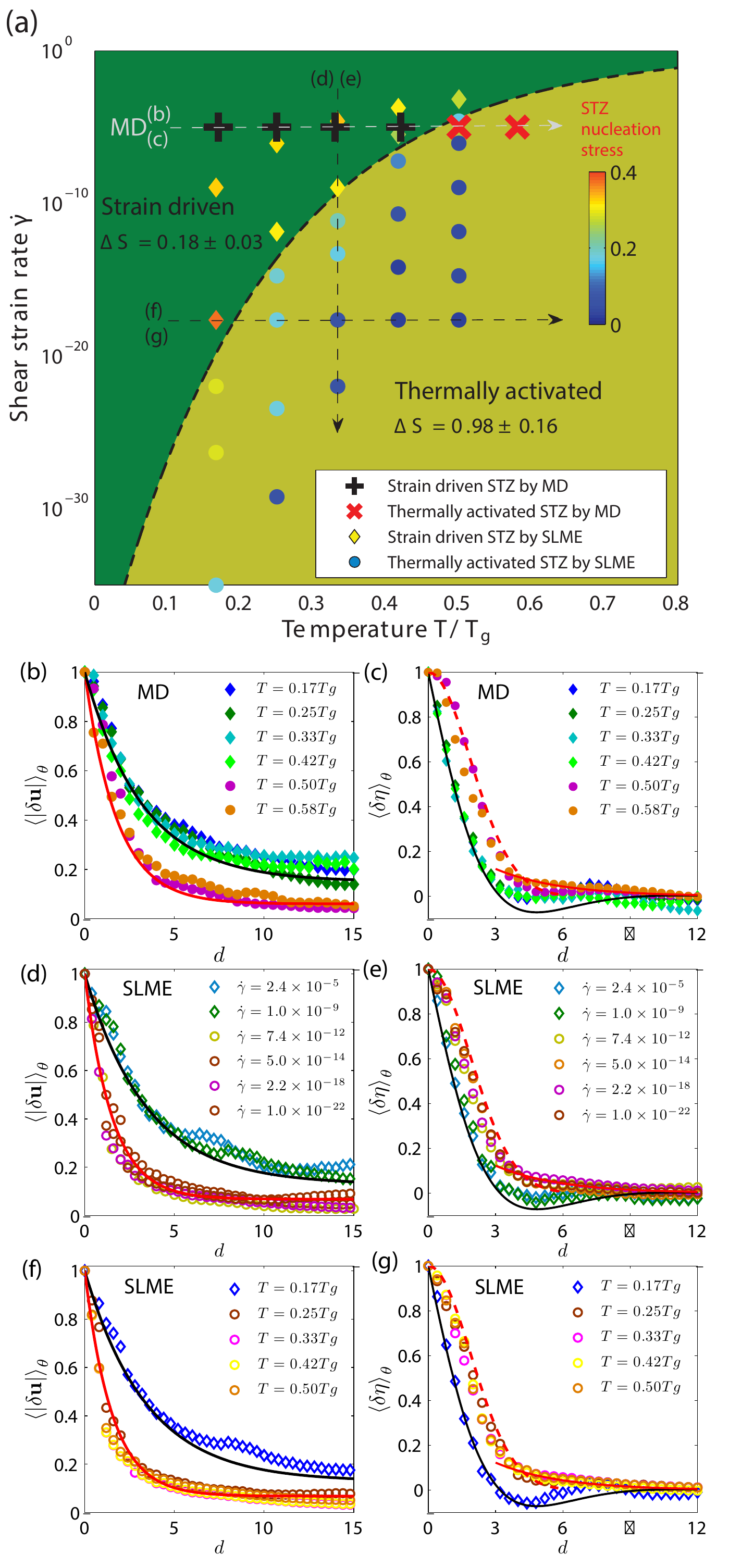}
\caption{ (Color online)(a) Summary of transition between strain-driven to thermally activated STZ nucleation as a function of strain rate and temperature, while also showing the dependence on the STZ nucleation stress, and the change in STZ area $\Delta S$ as computed using a Voronoi decomposition.  The black dashed line is the fitting function $\log_{10}(\dot{\gamma}) = -42.17\exp(-4.53T/T_g)$. Normalized  $\left<|\delta{\bf u}|\right>_{\theta}$ and $\left<\delta{\eta}\right>_{\theta}$ versus $d$ for various (b and c) temperatures at $\dot{\gamma}=1\times10^{-5}$ via classical MD simulations, (d)(e) strain rates at $T=0.33T_{g}$ by SLME, and (f)(g) temperatures at $\dot{\gamma}=2.2\times10^{-18}$ by SLME.  The black and red  lines in (b)-(g) correspond to the AQS and purely thermal results, respectively.}
\label{fig3} \end{center} \end{figure}

Having established the two distinct STZ types (strain driven and thermally activated) above, we now utilize the SLME algorithm coupled with shear deformation to study the coupled effects of temperature and strain rate on the characteristics of 2D STZs, where the results are summarized in Fig. \ref{fig3}.  Specifically, as shown in Fig. \ref{fig3}(a), there are effectively two regions, the thermally activated region for high $T$ and slow $\dot{\gamma}$ (yellow area), and the strain-driven region for low $T$ and high $\dot{\gamma}$ (green area). 

Before discussing the nature of the transition from strain-driven and thermally activated STZs at slow (experimentally relevant) strain rates that is summarized in Fig. \ref{fig3}(a), it is important to first establish as one means of validating the SLME approach that this universal transition over broad temperature and strain-rate ranges that is captured by the SLME approach can also be observed in actual MD simulations.  Therefore, we performed classical MD simulations at a strain rate of $\dot{\gamma}=1\times10^{-5}$, for a range of temperatures from 0.17$T_{g}$ to 0.58$T_{g}$.  As shown in Fig. \ref{fig3}(a), according to the MD simulations the STZ characteristics change from strain driven (black pluses) to thermally-activated (red crosses) around 0.5$T_{g}$, as shown in Fig. \ref{fig3}(a), matching the predictions of the SLME approach.  Furthermore, the sharp transition at about 0.5$T_{g}$ is characterized, as seen in Figs. \ref{fig3}(b) and (c), by transitions in both the displacement and the strain fields that were previously shown to be due to the transition in STZ characteristics from strain-driven to thermally activated.

Having established that MD and SLME agree for high strain rates, we now use the SLME approach to access those strain rates (i.e., ten or more orders of magnitude smaller than MD) that can be considered to be experimentally accessible.  In doing so, we find that at an experimentally relevant temperature $T=0.33T_{\rm g}$, Fig. \ref{fig3}(d) indicates that all the curves for $\dot{\gamma} \ge 1.0\times10^{-9}$ coincide with the AQS (and Eshelby) results, and all the curves for $\dot{\gamma} \le 7.4\times10^{-12}$ coincide with the purely thermal results.  This indicates that the characteristics of the STZ are strain dominated for the former and thermally assisted for the latter.  These results are important because the strain rate at which the transition from strain-dominated to thermally dominated STZ nucleation occurs corresponds to one that occurs at the strain rates that are experimentally relevant ($\dot{\gamma} =1\times 10^{-14}$), i.e. about ten orders of magnitude smaller than the MD strain rate ($\dot{\gamma} = 1\times10^{-5}$). This carries the important implication that if the STZ quadrupolar symmetry is broken at experimental strain rates, there may be an effect on the resulting shearband formation that occurs due to the STZ interactions that would be not be captured in artificially high strain rate MD simulations.

Thus, while Figs. \ref{fig3}(d) and (e) demonstrate the transition in STZ characteristics that occurs due to reducing the strain rate from MD to experimental, Figs. \ref{fig3}(f) and (g) demonstrate that the same transition in STZ characteristics can be achieved by keeping a constant strain rate, but increasing the temperature.  As a summary of all the results demonstrated in Figs. \ref{fig3}(b)-(g), Fig. \ref{fig3}(a) indicates that the transition from strain-dominated (Eshelby, or AQS) STZ nucleation to thermally dominated can be observed for strain rates ranging from MD ($\dot{\gamma}=1\times10^{-5}$) to experimental ($\dot{\gamma}=1\times10^{-15}$) by increasing the temperature, or by reducing the strain rate at constant temperature.  Moreover, the stress needed to nucleate the initial STZ decreases with increasing temperature, or equivalently with decreasing strain rate.  

It is also relevant to consider the effects of system size on the present results.  This is important because previous studies, such as done by Lerner and Procaccia ~\citet{lernerPRE2009}, have demonstrated that the yield stress, and the stress drop after yield, among other interesting quantities, are indeed system size dependent, where the system size refers to a system of $N$ atoms having periodic boundary conditions such that surface effects are not considered.  For our results in Fig. \ref{fig3}, the STZ transition mechanism we observed from strain driven to thermally assisted does not change.  However, what will change with the system size is the position of the transition curve in Fig. \ref{fig3}(a).

Before moving on, it is relevant to discuss here the effects of dimensionality on the results reported here, i.e., whether this sharp transition in STZ characteristics would be seen in three-dimensional (3D) amorphous solids.  Some skepticism as to whether such a finding would hold in three dimensional arises from recent results by ~\citet{olssonPRL2007},~\citet{guanPRL2010} and~\citet{langerPRE2012}, where stress-density, stress-temperature, and stress-viscosity scaling relations, respectively, were observed.  It is at present unclear whether the sharp transition in STZ characteristics that we have observed here in two dimensions would be observed in three dimensions, and is an important issue we will consider in future work.

\begin{figure} \begin{center} 
\includegraphics[scale=0.5]{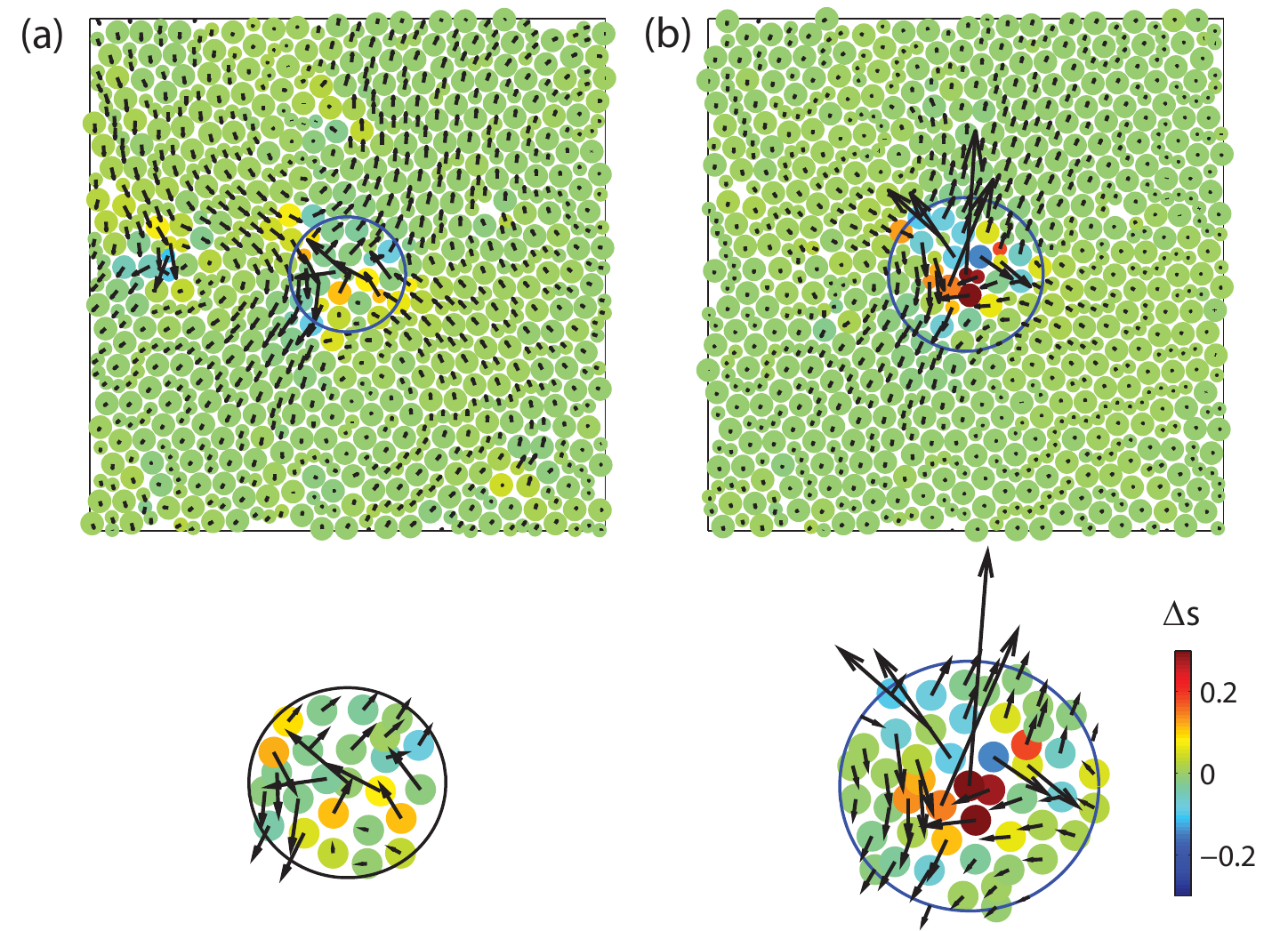}
\caption{(Color online) Change in Voronoi area $\Delta{S}$ of each atom between the undeformed configuration and STZ nucleation at $T=0.33T_{g}$ for strain rates of (a) $2.4\times10^{-5}$ and (b) $5.0\times10^{-14}$.}
\label{fig5} \end{center} \end{figure} 

We address the transition in the deformation mechanism inside the STZ that occurs in transitioning from strain-induced to thermally-activated STZ formation.  Figure \ref{fig5} shows the change in Voronoi area $\Delta{S}$ within the STZ core at $T=0.33T_{g}$ for shear strain rates of $2.4\times10^{-5}$ (left) and $5.0\times10^{-14}$ (right).  We find that the Voronoi area for the STZ at the faster, MD-relevant strain rate of $2.4\times10^{-5}$ is about 0.21, whereas a much larger increase of 0.93 is found for the experimental strain rate of $5.0\times10^{-14}$.  These values are representative of the average change in Voronoi area that we calculate across the range of different strain rates and temperatures we considered as shown in Fig. \ref{fig3}(a), where the average $\Delta{S}$ for strain-driven STZ nucleation is about 0.18, while the average $\Delta{S}$ for thermally activated STZ nucleation is about 0.98.  Furthermore, the significantly faster decay of the nonaffine displacement field at the experimental strain rate that was previously quantified in Fig. \ref{fig3} is seen clearly in comparing Fig. \ref{fig5}(b) to Fig. \ref{fig5}(a).  

Most interestingly, the deformation inside the STZ changes from shear dominated at elevated strain rates, which corresponds well to the relatively small change in STZ Voronoi area as illustrated in Fig. \ref{fig5}(a), to tensile once the deformation is thermally activated as shown in Fig. \ref{fig5}(b), which also agrees well with the larger change in STZ Voronoi area.  Furthermore, we can observe that the feature that coincides with the shear-to-tensile deformation change inside the STZ is the breaking of the quadrupolar symmetry that is seen in Fig. \ref{fig5}(a), where in Fig. \ref{fig5}(b) the compressive portion of the quadrupolar deformation is significantly reduced, whereas the tensile portion remains largely intact.  This suppression of the compressive portion of the quadrupole renders the STZ deformation largely tensile at slower strain rates, resulting in the larger STZ area changes, whereas the largely shear-dominated STZ deformation results in very small area changes, which is consistent with a shear (volume-preserving) deformation mechanism.

As a final, but important, comment, we note that no comparison of the strain-driven or thermally activated STZs to experimental results have been made in this work.  This fact is not particular to this simulation study, but is in fact a general theme of all atomistic simulation studies of STZs in amorphous solid due to the fact that the atomic scale structure of an individual STZ has yet to be resolved experimentally~\cite{schuhAM2007,chengPMS2011}.  In contrast, current experimental studies are able to back out the STZ size by using activation energy arguments~\cite{panPNAS2008,harmonPRL2007}.  Thus, it is hoped that the theoretical results of STZ characteristics obtained in this work may prove beneficial to theorists and experimentalists alike in the future.

\section{Conclusion}

In conclusion, we have coupled a recent PES exploration technique with shear deformation to study the characteristics of STZs in a two-dimensional binary Lennard-Jones amorphous solid.  Specifically, we report a transition in the STZ characteristics where the transition can occur either by increasing the temperature or by decreasing the strain rate.  The transition occurs between STZs that have characteristics identical to those that are found in purely strain-driven, AQS calculations, and those that emerge from a purely thermally activated state.  The specific changes in STZ characteristics that occur in moving from the strain-driven to the thermally activated STZ regime include a 33\% increase in STZ size, faster spatial decay of the displacement field, a change in deformation mechanism inside the STZ from shear to tension, a reduction in the stress needed to nucleate the first STZ, and finally a notable loss in the characteristic quadrupolar symmetry of the surrounding elastic matrix that has previously been seen in athermal, quasistatic shear studies of STZs.

\section{ACKNOWLEDGMENT}
All authors acknowledge the support of the NSF through Grant No. CMMI-1234183. X.L. also acknowledges the support of Nenter $\&$ Co., Inc. and NSF-XSEDE through Grant
No. DMR-0900073.

\bibliography{biball}

\begin{thebibliography}{35}
\expandafter\ifx\csname natexlab\endcsname\relax\def\natexlab#1{#1}\fi
\expandafter\ifx\csname bibnamefont\endcsname\relax
  \def\bibnamefont#1{#1}\fi
\expandafter\ifx\csname bibfnamefont\endcsname\relax
  \def\bibfnamefont#1{#1}\fi
\expandafter\ifx\csname citenamefont\endcsname\relax
  \def\citenamefont#1{#1}\fi
\expandafter\ifx\csname url\endcsname\relax
  \def\url#1{\texttt{#1}}\fi
\expandafter\ifx\csname urlprefix\endcsname\relax\def\urlprefix{URL }\fi
\providecommand{\bibinfo}[2]{#2}
\providecommand{\eprint}[2][]{\url{#2}}

\bibitem[{\citenamefont{Schuh et~al.}(2007)\citenamefont{Schuh, Hufnagel, and
  Ramamurty}}]{schuhAM2007}
\bibinfo{author}{\bibfnamefont{C.~A.} \bibnamefont{Schuh}},
  \bibinfo{author}{\bibfnamefont{T.~C.} \bibnamefont{Hufnagel}},
  \bibnamefont{and}
  \bibinfo{author}{\bibfnamefont{U.}~\bibnamefont{Ramamurty}},
  \bibinfo{journal}{Acta Materialia} \textbf{\bibinfo{volume}{55}},
  \bibinfo{pages}{4067} (\bibinfo{year}{2007}).

\bibitem[{\citenamefont{Cheng and Ma}(2011{\natexlab{a}})}]{chengPMS2011}
\bibinfo{author}{\bibfnamefont{Y.~Q.} \bibnamefont{Cheng}} \bibnamefont{and}
  \bibinfo{author}{\bibfnamefont{E.}~\bibnamefont{Ma}},
  \bibinfo{journal}{Progress in Materials Science}
  \textbf{\bibinfo{volume}{56}}, \bibinfo{pages}{379}
  (\bibinfo{year}{2011}{\natexlab{a}}).

\bibitem[{\citenamefont{Argon}(1979)}]{argonAM1979}
\bibinfo{author}{\bibfnamefont{A.~S.} \bibnamefont{Argon}},
  \bibinfo{journal}{Acta Metallurgica} \textbf{\bibinfo{volume}{27}},
  \bibinfo{pages}{47} (\bibinfo{year}{1979}).

\bibitem[{\citenamefont{Maloney and Lemaitre}(2006)}]{maloneyPRE2006}
\bibinfo{author}{\bibfnamefont{C.~E.} \bibnamefont{Maloney}} \bibnamefont{and}
  \bibinfo{author}{\bibfnamefont{A.}~\bibnamefont{Lemaitre}},
  \bibinfo{journal}{Physical Review E} \textbf{\bibinfo{volume}{74}},
  \bibinfo{pages}{016118} (\bibinfo{year}{2006}).

\bibitem[{\citenamefont{Tsamados et~al.}(2009)\citenamefont{Tsamados, Tanguy,
  Goldenberg, and Barrat}}]{tsamadosPRE2009}
\bibinfo{author}{\bibfnamefont{M.}~\bibnamefont{Tsamados}},
  \bibinfo{author}{\bibfnamefont{A.}~\bibnamefont{Tanguy}},
  \bibinfo{author}{\bibfnamefont{C.}~\bibnamefont{Goldenberg}},
  \bibnamefont{and} \bibinfo{author}{\bibfnamefont{J.-L.}
  \bibnamefont{Barrat}}, \bibinfo{journal}{Physical Review E}
  \textbf{\bibinfo{volume}{80}}, \bibinfo{pages}{026112}
  (\bibinfo{year}{2009}).

\bibitem[{\citenamefont{Tanguy et~al.}(2006)\citenamefont{Tanguy, Leonforte,
  and Barrat}}]{tanguyEPJE2006}
\bibinfo{author}{\bibfnamefont{A.}~\bibnamefont{Tanguy}},
  \bibinfo{author}{\bibfnamefont{F.}~\bibnamefont{Leonforte}},
  \bibnamefont{and} \bibinfo{author}{\bibfnamefont{J.-L.}
  \bibnamefont{Barrat}}, \bibinfo{journal}{The European Physical Journal E}
  \textbf{\bibinfo{volume}{20}}, \bibinfo{pages}{355} (\bibinfo{year}{2006}).

\bibitem[{\citenamefont{Lemaitre and Caroli}(2007)}]{lemaitrePRE2007}
\bibinfo{author}{\bibfnamefont{A.}~\bibnamefont{Lemaitre}} \bibnamefont{and}
  \bibinfo{author}{\bibfnamefont{C.}~\bibnamefont{Caroli}},
  \bibinfo{journal}{Phys. Rev. E} \textbf{\bibinfo{volume}{76}},
  \bibinfo{pages}{036104} (\bibinfo{year}{2007}).

\bibitem[{\citenamefont{Zink et~al.}(2006)\citenamefont{Zink, Samwer, Johnson,
  and Mayr}}]{zinkPRB2006}
\bibinfo{author}{\bibfnamefont{M.}~\bibnamefont{Zink}},
  \bibinfo{author}{\bibfnamefont{K.}~\bibnamefont{Samwer}},
  \bibinfo{author}{\bibfnamefont{W.~L.} \bibnamefont{Johnson}},
  \bibnamefont{and} \bibinfo{author}{\bibfnamefont{S.~G.} \bibnamefont{Mayr}},
  \bibinfo{journal}{Physical Review B} \textbf{\bibinfo{volume}{73}},
  \bibinfo{pages}{172203} (\bibinfo{year}{2006}).

\bibitem[{\citenamefont{Falk and Langer}(1998)}]{falkPRE1998}
\bibinfo{author}{\bibfnamefont{M.~L.} \bibnamefont{Falk}} \bibnamefont{and}
  \bibinfo{author}{\bibfnamefont{J.~S.} \bibnamefont{Langer}},
  \bibinfo{journal}{Physical Review E} \textbf{\bibinfo{volume}{57}},
  \bibinfo{pages}{7192} (\bibinfo{year}{1998}).

\bibitem[{\citenamefont{Rodney and Schuh}(2009{\natexlab{a}})}]{rodneyPRL2009}
\bibinfo{author}{\bibfnamefont{D.}~\bibnamefont{Rodney}} \bibnamefont{and}
  \bibinfo{author}{\bibfnamefont{C.~A.} \bibnamefont{Schuh}},
  \bibinfo{journal}{Physical Review Letters} \textbf{\bibinfo{volume}{102}},
  \bibinfo{pages}{235503} (\bibinfo{year}{2009}{\natexlab{a}}).

\bibitem[{\citenamefont{Rodney and Schuh}(2009{\natexlab{b}})}]{rodneyPRB2009}
\bibinfo{author}{\bibfnamefont{D.}~\bibnamefont{Rodney}} \bibnamefont{and}
  \bibinfo{author}{\bibfnamefont{C.~A.} \bibnamefont{Schuh}},
  \bibinfo{journal}{Physical Review B} \textbf{\bibinfo{volume}{80}},
  \bibinfo{pages}{184203} (\bibinfo{year}{2009}{\natexlab{b}}).

\bibitem[{\citenamefont{Eshelby}(1957)}]{eshelbyPTS1957}
\bibinfo{author}{\bibfnamefont{J.~D.} \bibnamefont{Eshelby}},
  \bibinfo{journal}{Proceedings of the Royal Society of London. Series A.
  Mathematical and Physical Sciences} \textbf{\bibinfo{volume}{241}},
  \bibinfo{pages}{376} (\bibinfo{year}{1957}).

\bibitem[{\citenamefont{Dasgupta et~al.}(2012)\citenamefont{Dasgupta,
  Hentschel, and Procaccia}}]{dasguptaPRL2012}
\bibinfo{author}{\bibfnamefont{R.}~\bibnamefont{Dasgupta}},
  \bibinfo{author}{\bibfnamefont{H.~G.~E.} \bibnamefont{Hentschel}},
  \bibnamefont{and}
  \bibinfo{author}{\bibfnamefont{I.}~\bibnamefont{Procaccia}},
  \bibinfo{journal}{Physical Review Letters} \textbf{\bibinfo{volume}{109}},
  \bibinfo{pages}{255502} (\bibinfo{year}{2012}).

\bibitem[{\citenamefont{Cao et~al.}(2012)\citenamefont{Cao, Li, Heugle, Park,
  and Lin}}]{caoPRE2012}
\bibinfo{author}{\bibfnamefont{P.}~\bibnamefont{Cao}},
  \bibinfo{author}{\bibfnamefont{M.}~\bibnamefont{Li}},
  \bibinfo{author}{\bibfnamefont{R.~J.} \bibnamefont{Heugle}},
  \bibinfo{author}{\bibfnamefont{H.~S.} \bibnamefont{Park}}, \bibnamefont{and}
  \bibinfo{author}{\bibfnamefont{X.}~\bibnamefont{Lin}},
  \bibinfo{journal}{Physical Review E} \textbf{\bibinfo{volume}{86}},
  \bibinfo{pages}{016710} (\bibinfo{year}{2012}).

\bibitem[{\citenamefont{Kushima
  et~al.}(2009{\natexlab{a}})\citenamefont{Kushima, Lin, Li, Eapen, Mauro,
  Qian, Diep, and Yip}}]{kushimaJCP2009}
\bibinfo{author}{\bibfnamefont{A.}~\bibnamefont{Kushima}},
  \bibinfo{author}{\bibfnamefont{X.}~\bibnamefont{Lin}},
  \bibinfo{author}{\bibfnamefont{J.}~\bibnamefont{Li}},
  \bibinfo{author}{\bibfnamefont{J.}~\bibnamefont{Eapen}},
  \bibinfo{author}{\bibfnamefont{J.~C.} \bibnamefont{Mauro}},
  \bibinfo{author}{\bibfnamefont{X.}~\bibnamefont{Qian}},
  \bibinfo{author}{\bibfnamefont{P.}~\bibnamefont{Diep}}, \bibnamefont{and}
  \bibinfo{author}{\bibfnamefont{S.}~\bibnamefont{Yip}},
  \bibinfo{journal}{Journal of Chemical Physics}
  \textbf{\bibinfo{volume}{130}}, \bibinfo{pages}{224504}
  (\bibinfo{year}{2009}{\natexlab{a}}).

\bibitem[{\citenamefont{Kushima
  et~al.}(2009{\natexlab{b}})\citenamefont{Kushima, Lin, Li, Qian, Eapen,
  Mauro, Diep, and Yip}}]{kushimaJCP2009_2}
\bibinfo{author}{\bibfnamefont{A.}~\bibnamefont{Kushima}},
  \bibinfo{author}{\bibfnamefont{X.}~\bibnamefont{Lin}},
  \bibinfo{author}{\bibfnamefont{J.}~\bibnamefont{Li}},
  \bibinfo{author}{\bibfnamefont{X.}~\bibnamefont{Qian}},
  \bibinfo{author}{\bibfnamefont{J.}~\bibnamefont{Eapen}},
  \bibinfo{author}{\bibfnamefont{J.~C.} \bibnamefont{Mauro}},
  \bibinfo{author}{\bibfnamefont{P.}~\bibnamefont{Diep}}, \bibnamefont{and}
  \bibinfo{author}{\bibfnamefont{S.}~\bibnamefont{Yip}}, \bibinfo{journal}{The
  Journal of chemical physics} \textbf{\bibinfo{volume}{131}},
  \bibinfo{pages}{164505} (\bibinfo{year}{2009}{\natexlab{b}}).

\bibitem[{\citenamefont{Laio and Parrinello}(2002)}]{laioPNAS2002}
\bibinfo{author}{\bibfnamefont{A.}~\bibnamefont{Laio}} \bibnamefont{and}
  \bibinfo{author}{\bibfnamefont{M.}~\bibnamefont{Parrinello}},
  \bibinfo{journal}{Proceedings of the National Academy of Science}
  \textbf{\bibinfo{volume}{99}}, \bibinfo{pages}{12562} (\bibinfo{year}{2002}).

\bibitem[{\citenamefont{Voter}(1997)}]{voterPRL1997}
\bibinfo{author}{\bibfnamefont{A.~F.} \bibnamefont{Voter}},
  \bibinfo{journal}{Physical Review Letters} \textbf{\bibinfo{volume}{78}},
  \bibinfo{pages}{3908} (\bibinfo{year}{1997}).

\bibitem[{\citenamefont{Henkelman and Jonsson}(1999)}]{henkelmanJCP1999}
\bibinfo{author}{\bibfnamefont{G.}~\bibnamefont{Henkelman}} \bibnamefont{and}
  \bibinfo{author}{\bibfnamefont{H.}~\bibnamefont{Jonsson}},
  \bibinfo{journal}{Journal of Chemical Physics}
  \textbf{\bibinfo{volume}{111}}, \bibinfo{pages}{7010} (\bibinfo{year}{1999}).

\bibitem[{\citenamefont{Zhu et~al.}(2008)\citenamefont{Zhu, Li, Samanta, Leach,
  and Gall}}]{zhuPRL2008}
\bibinfo{author}{\bibfnamefont{T.}~\bibnamefont{Zhu}},
  \bibinfo{author}{\bibfnamefont{J.}~\bibnamefont{Li}},
  \bibinfo{author}{\bibfnamefont{A.}~\bibnamefont{Samanta}},
  \bibinfo{author}{\bibfnamefont{A.}~\bibnamefont{Leach}}, \bibnamefont{and}
  \bibinfo{author}{\bibfnamefont{K.}~\bibnamefont{Gall}},
  \bibinfo{journal}{Physical Review Letters} \textbf{\bibinfo{volume}{100}},
  \bibinfo{pages}{025502} (\bibinfo{year}{2008}).

\bibitem[{\citenamefont{Park et~al.}(2006)\citenamefont{Park, Gall, and
  Zimmerman}}]{parkJMPS2006}
\bibinfo{author}{\bibfnamefont{H.~S.} \bibnamefont{Park}},
  \bibinfo{author}{\bibfnamefont{K.}~\bibnamefont{Gall}}, \bibnamefont{and}
  \bibinfo{author}{\bibfnamefont{J.~A.} \bibnamefont{Zimmerman}},
  \bibinfo{journal}{Journal of the Mechanics and Physics of Solids}
  \textbf{\bibinfo{volume}{54}}, \bibinfo{pages}{1862} (\bibinfo{year}{2006}).

\bibitem[{\citenamefont{Ryu et~al.}(2011)\citenamefont{Ryu, Kang, and
  Cai}}]{ryuPNAS2011}
\bibinfo{author}{\bibfnamefont{S.}~\bibnamefont{Ryu}},
  \bibinfo{author}{\bibfnamefont{K.}~\bibnamefont{Kang}}, \bibnamefont{and}
  \bibinfo{author}{\bibfnamefont{W.}~\bibnamefont{Cai}},
  \bibinfo{journal}{Proceedings of the National Academy of Science}
  \textbf{\bibinfo{volume}{108}}, \bibinfo{pages}{5174} (\bibinfo{year}{2011}).

\bibitem[{\citenamefont{Li et~al.}(2011)\citenamefont{Li, Kushima, Eapen, Lin,
  Qian, Mauro, Diep, and Yip}}]{liPO2011}
\bibinfo{author}{\bibfnamefont{J.}~\bibnamefont{Li}},
  \bibinfo{author}{\bibfnamefont{A.}~\bibnamefont{Kushima}},
  \bibinfo{author}{\bibfnamefont{J.}~\bibnamefont{Eapen}},
  \bibinfo{author}{\bibfnamefont{X.}~\bibnamefont{Lin}},
  \bibinfo{author}{\bibfnamefont{X.}~\bibnamefont{Qian}},
  \bibinfo{author}{\bibfnamefont{J.}~\bibnamefont{Mauro}},
  \bibinfo{author}{\bibfnamefont{P.}~\bibnamefont{Diep}}, \bibnamefont{and}
  \bibinfo{author}{\bibfnamefont{S.}~\bibnamefont{Yip}}, \bibinfo{journal}{PLoS
  One} \textbf{\bibinfo{volume}{6}}, \bibinfo{pages}{e17909}
  (\bibinfo{year}{2011}).

\bibitem[{\citenamefont{Johnson and Samwer}(2005)}]{johnsonPRL2005}
\bibinfo{author}{\bibfnamefont{W.~L.} \bibnamefont{Johnson}} \bibnamefont{and}
  \bibinfo{author}{\bibfnamefont{K.}~\bibnamefont{Samwer}},
  \bibinfo{journal}{Physical Review Letters} \textbf{\bibinfo{volume}{95}},
  \bibinfo{pages}{195501} (\bibinfo{year}{2005}).

\bibitem[{\citenamefont{Cheng and Ma}(2011{\natexlab{b}})}]{chengAM2011}
\bibinfo{author}{\bibfnamefont{Y.~Q.} \bibnamefont{Cheng}} \bibnamefont{and}
  \bibinfo{author}{\bibfnamefont{E.}~\bibnamefont{Ma}}, \bibinfo{journal}{Acta
  Materialia} \textbf{\bibinfo{volume}{59}}, \bibinfo{pages}{1800}
  (\bibinfo{year}{2011}{\natexlab{b}}).

\bibitem[{\citenamefont{Wales}(2003)}]{wales2003}
\bibinfo{author}{\bibfnamefont{D.~J.} \bibnamefont{Wales}},
  \emph{\bibinfo{title}{Energy landscapes: With applications to clusters,
  biomolecules and glasses}} (\bibinfo{publisher}{Cambridge University Press},
  \bibinfo{year}{2003}), ISBN \bibinfo{isbn}{0-521-814157-4}.

\bibitem[{\citenamefont{Allen and Tildesley}(1987)}]{allen1987}
\bibinfo{author}{\bibfnamefont{M.~P.} \bibnamefont{Allen}} \bibnamefont{and}
  \bibinfo{author}{\bibfnamefont{D.~J.} \bibnamefont{Tildesley}},
  \emph{\bibinfo{title}{Computer Simulation of Liquids}}
  (\bibinfo{publisher}{Oxford University Press}, \bibinfo{year}{1987}).

\bibitem[{\citenamefont{Yoshimoto et~al.}(2004)\citenamefont{Yoshimoto, Jain,
  Workum, Nealey, and de~Pablo}}]{YoshimotoPRL2004}
\bibinfo{author}{\bibfnamefont{K.}~\bibnamefont{Yoshimoto}},
  \bibinfo{author}{\bibfnamefont{T.~S.} \bibnamefont{Jain}},
  \bibinfo{author}{\bibfnamefont{K.~V.} \bibnamefont{Workum}},
  \bibinfo{author}{\bibfnamefont{P.~F.} \bibnamefont{Nealey}},
  \bibnamefont{and} \bibinfo{author}{\bibfnamefont{J.~J.}
  \bibnamefont{de~Pablo}}, \bibinfo{journal}{Physical Review Letters}
  \textbf{\bibinfo{volume}{93}}, \bibinfo{pages}{175501}
  (\bibinfo{year}{2004}).

\bibitem[{\citenamefont{Lutsko}(1988)}]{lutskoJAP1988}
\bibinfo{author}{\bibfnamefont{J.}~\bibnamefont{Lutsko}},
  \bibinfo{journal}{Journal of applied physics} \textbf{\bibinfo{volume}{64}},
  \bibinfo{pages}{1152} (\bibinfo{year}{1988}).

\bibitem[{\citenamefont{Lerner and Procaccia}(2009)}]{lernerPRE2009}
\bibinfo{author}{\bibfnamefont{E.}~\bibnamefont{Lerner}} \bibnamefont{and}
  \bibinfo{author}{\bibfnamefont{I.}~\bibnamefont{Procaccia}},
  \bibinfo{journal}{Phys. Rev. E} \textbf{\bibinfo{volume}{79}},
  \bibinfo{pages}{066109} (\bibinfo{year}{2009}),
  \urlprefix\url{http://link.aps.org/doi/10.1103/PhysRevE.79.066109}.

\bibitem[{\citenamefont{Olsson and Teitel}(2007)}]{olssonPRL2007}
\bibinfo{author}{\bibfnamefont{P.}~\bibnamefont{Olsson}} \bibnamefont{and}
  \bibinfo{author}{\bibfnamefont{S.}~\bibnamefont{Teitel}},
  \bibinfo{journal}{Physical Review Letters} \textbf{\bibinfo{volume}{99}},
  \bibinfo{pages}{178001} (\bibinfo{year}{2007}).

\bibitem[{\citenamefont{Guan et~al.}(2010)\citenamefont{Guan, Chen, and
  Egami}}]{guanPRL2010}
\bibinfo{author}{\bibfnamefont{P.}~\bibnamefont{Guan}},
  \bibinfo{author}{\bibfnamefont{M.}~\bibnamefont{Chen}}, \bibnamefont{and}
  \bibinfo{author}{\bibfnamefont{T.}~\bibnamefont{Egami}},
  \bibinfo{journal}{Physical Review Letters} \textbf{\bibinfo{volume}{104}},
  \bibinfo{pages}{205701} (\bibinfo{year}{2010}).

\bibitem[{\citenamefont{Langer and Egami}(2012)}]{langerPRE2012}
\bibinfo{author}{\bibfnamefont{J.~S.} \bibnamefont{Langer}} \bibnamefont{and}
  \bibinfo{author}{\bibfnamefont{T.}~\bibnamefont{Egami}},
  \bibinfo{journal}{Phys. Rev. E} \textbf{\bibinfo{volume}{86}},
  \bibinfo{pages}{011502} (\bibinfo{year}{2012}).

\bibitem[{\citenamefont{Pan et~al.}(2008)\citenamefont{Pan, Inoue, Sakurai, and
  Chen}}]{panPNAS2008}
\bibinfo{author}{\bibfnamefont{D.}~\bibnamefont{Pan}},
  \bibinfo{author}{\bibfnamefont{A.}~\bibnamefont{Inoue}},
  \bibinfo{author}{\bibfnamefont{T.}~\bibnamefont{Sakurai}}, \bibnamefont{and}
  \bibinfo{author}{\bibfnamefont{M.~W.} \bibnamefont{Chen}},
  \bibinfo{journal}{Proceedings of the National Academy of Science}
  \textbf{\bibinfo{volume}{105}}, \bibinfo{pages}{14769}
  (\bibinfo{year}{2008}).

\bibitem[{\citenamefont{Harmon et~al.}(2007)\citenamefont{Harmon, Demetriou,
  Johnson, and Samwer}}]{harmonPRL2007}
\bibinfo{author}{\bibfnamefont{J.~S.} \bibnamefont{Harmon}},
  \bibinfo{author}{\bibfnamefont{M.~D.} \bibnamefont{Demetriou}},
  \bibinfo{author}{\bibfnamefont{W.~L.} \bibnamefont{Johnson}},
  \bibnamefont{and} \bibinfo{author}{\bibfnamefont{K.}~\bibnamefont{Samwer}},
  \bibinfo{journal}{Physical Review Letters} \textbf{\bibinfo{volume}{99}},
  \bibinfo{pages}{135502} (\bibinfo{year}{2007}).

\end{thebibliography}

\end{document}